\documentclass[]{spie}  %>>> use for US letter paper
\usepackage{amsmath,amsfonts,amssymb}
\usepackage{graphicx}
\usepackage[colorlinks=true, allcolors=blue]{hyperref}

\title{CHARA/MIRC-X - a high-sensitive six telescope interferometric imager concept, commissioning, and early science}

\author[a,b,*]{Narsireddy Anugu}
\author[c]{Jean-Baptiste Le~Bouquin}
\author[d]{John D. Monnier}
\author[b]{Stefan Kraus}
\author[e]{Gail Schaefer}
\author[d]{Benjamin R. Setterholm}
\author[b]{Claire L Davies}
\author[d]{Tyler Gardner}
\author[b]{Aaron Labdon}
\author[d]{Cyprien Lanthermann}
\author[d]{Jacob Ennis}
\author[e]{Theo ten Brummelaar}
\author[e]{Judit Sturmann}
\author[e]{Matt Anderson}
\author[e]{Chris Farrington}
\author[e]{Norm Vargas}
\author[e]{Olli Majoinen}

\affil[a]{Steward Observatory, Department of Astronomy, University of Arizona, Tucson, USA}
\affil[b]{School of Physics and Astronomy, University of Exeter,  Exeter, Stocker Road, EX4 4QL, UK}
\affil[c]{Institut de Planetologie et d’Astrophysique de Grenoble, Grenoble 38058, France}
\affil[d]{University of Michigan, Ann Arbor, MI 48109, USA}
\affil[e]{CHARA Array, Georgia State University, Atlanta, GA 30302, USA}

\authorinfo{* Steward Observatory Fellow in Astronomical Instrumentation and
Technology. \\
E-mail: \href{nanugu@arizona.edu}{nanugu@arizona.edu}}

\begin{document} 
\maketitle

\begin{abstract}
MIRC-X is a six telescope beam combiner at the CHARA array that works in J and H wavelength bands and provides an angular resolution equivalent to a $B$=331m diameter telescope.  The legacy MIRC combiner has delivered outstanding results in the fields of stellar astrophysics and binaries. However, we required higher sensitivity to make ambitious scientific measurements of faint targets such as young stellar objects, binary systems with exoplanets, and active galactic nuclei. For that purpose, MIRC-X is built and is offered to the community since mid-2017.  MIRC-X has demonstrated up to two magnitudes of improved faint magnitude sensitivity with the best-case H $<= 8$. Here we present a review of the instrument and present early science results, and highlight some of our ongoing science programs.
\end{abstract}

% Include a list of keywords after the abstract 

\section{INTRODUCTION}\label{sec:intro}  
High-angular resolution observations of disks around young stellar objects (YSOs) such as Herbig Ae/Be and T~Tauri stars are vital to advance our understanding of the star and planet formation process \cite{Dullemond2010ARA&A..48..205D}. SPHERE\cite{Beuzit2019}, GPI\cite{Macintosh2014}, and ALMA\cite{ALMA2015} are among the most powerful facilities for making high-angular observations, and they have contributed immensely in understanding the YSO systems\cite{Andrews2018ApJ...869L..41A, Avenhaus2018ApJ...863...44A}. These instruments have probed outer disk features of YSOs (i.e., approximately 20 to 500 astronomical units, au at 100 parsecs) in great detail at near-infrared and sub-millimeter wavelengths. Despite these successes, the aforementioned facilities lack spatial resolutions to cover the inner-most regions of YSO disks close to the dust sublimation rim, ranging the size of a few au, where dynamical disk evolution, terrestrial planet formation, and planet migration most likely takes place. Optical interferometric instruments at Very Large Telescope Interferometer (VLTI) such as PIONIER\cite{Lebouquin2011}, GRAVITY\cite{GRAVITY2017}, and MATISSE\cite{Lopez2014} cover the intermediate spatial scales of the disks\cite{Lazareff2017A&A...599A..85L,GravityCollaboration2019A&A...632A..53G}. However, they are limited by the availability of only four telescopes (poor uv-coverage) and relatively short baselines of the VLTI array. Also, the VLTI instruments are located in the Southern hemisphere with limited access to Northern targets.

We built the Michigan InfraRed Combiner - eXeter (MIRC-X)\cite{Anugu2020AJ....160..158A} instrument aiming to fill this gap. It is installed at the Center for High Angular Resolution Array\cite{tenBrummelaar2005, Schaefer2020}, located on Mt.\ Wilson, California, leveraging the six available telescopes and largest optical baselines in the world (up to $331$\,m, more than $2.5$ times longer than the maximum VLTI baselines). The detailed science cases of MIRC-X are described else where\cite{Anugu2020AJ....160..158A, Kraus2018}. MIRC-X is the evolution of the MIRC instrument\cite{Monnier2006, Monnier2010, Che2010}, which achieved landmark results in stellar astrophysics, for instance, imaging the fireball expansion phase of a nova explosion\cite{Schaefer2014}, the transit of an eclipsing binary system\cite{Kloppenborg2010} and surfaces and spots of other stars\cite{Monnier2007, Roettenbacher2016}. However, to achieve ambitious scientific measurements of faint YSOs MIRC was upgraded to MIRC-X. MIRC-X uses a sub-electron readout noise and fast-frame rate (3500Hz) C-RED ONE camera\cite{Gach2016} leveraging a revolutionary HgCdTe Electron Avalanche Photodiode Array (eAPD) technology\cite{Finger2014}. Additionally, MIRC-X will be utilized as a fringe tracker for the upcoming K-band MYSTIC\cite{Monnier2018} and the R-band SPICA\cite{Pannetier2020} instruments and takes simultaneous observations in J and H-bands.

MIRC-X is built with funding from the European Research Council (PI Prof. Stefan Kraus) and the National Science Foundation (PI Prof. John Monnier). MIRC-X was built in two phases. In Phase I: we upgraded the former MIRC camera with an ultra-low noise and fast frame rate, C-RED ONE, in place of the old and noisy PICNIC camera\cite{Anugu2018, Kraus2018}. Phase II, advanced the optics of the MIRC beam-train which were optimized for sensitivity\cite{Anugu2020AJ....160..158A}.  In this paper, we will summarize instrument capabilities and outline early on-sky science results.

\section{MIRC-X}
\subsection{Overview}
MIRC-X science cases, technical design, optics, software, development, and on-sky commissioning results are described in detail in Anugu et al. 2020\cite{Anugu2020AJ....160..158A}.  Although MIRC-X is an upgrade project, it is redesigned and rebuilt entirely, leveraging advancements in the detector technology and optics since the installation of the legacy MIRC instrument and exploiting the expertise of building instruments MIRC, PIONIER, and GRAVITY. A few highlights of MIRC-X compared to MIRC are:

\begin{itemize}
    \item \textbf{Sensitivity} is improved up to two magnitudes with the usage of state-of-the-art eAPD technology-based C-RED ONE camera.
    \item \textbf{High throughput photometric channels} The MIRC-X photometric channels are redesigned to have higher throughput and are more robust to alignment drifts.
    \item \textbf{Additional J-band observing mode} is commissioned with a wavelength dispersion control to enable simultaneous broad-band J+H observations. The J-band is mostly an unexplored window in long-baseline optical interferometry science.
    \item \textbf{Additional polar-interferometry observing mode} is made possible with new half-wave plates for each beam (modulator/retarder) and a Wollaston prism (analyzer) to measure linear Stokes parameters. An active polarization birefringence control is installed to maximize the visibility contrast. The polar-interferometry mode is the most untapped observing mode in the long-baseline optical interferometry science. 
    \item \textbf{Co-phase operation with MYSTIC and SPICA} Added differential delay lines and upgraded software for the coordinated observations with MYSTIC\cite{Monnier2018} and SPICA\cite{Pannetier2020}. MIRC-X will also act as a fringe tracker for SPICA and MYSTIC while simultaneously recording data in H-band. 
    \item \textbf{Six Telescope Simulator (STS)}  is installed for calibrating MIRC-X. STS proved to be very successful since its installation as an alignment check instrument for MIRC-X, and it is also used for science observations to calibrate the closure phases when multiple calibrator objects were unavailable\cite{Gardner2020}. 
    \item \textbf{Remote observing mode} is enabled with the upgrade of new MIRC-X control software\cite{Anugu2020AJ....160..158A}, which follows the CHARA compliant architecture instead of former Python-based GUIs of MIRC. In this remote observing mode, all the real-time tasks are executed by the servers that run on the on-site instrument computer. Observing clients (GUIs) run on a remote machine communicate to the server via Secure Shell tunneling using the remote SSH port forwarding protocol.
\end{itemize}

See Table~\ref{tab:MIRC-X_specs} for the summarized MIRC-X specifications.

\begin{table}[ht]
\caption{Summary of MIRC-X specifications\cite{Anugu2020AJ....160..158A}.}
\label{tab:MIRC-X_specs}
\begin{center}       
\begin{tabular}{ll}
\hline
Parameter &  Specification\\
\hline
Wavelength of operation & J+H wavelength bands, i.e., $1.1-1.8 ~\mu$m \\
Detector & C-RED ONE (0.3e-/px readout noise,   3500 frame rate) \\
Active differential delay line control & To co-phase operation with MYSTIC and SPICA\\
Active wavelength dispersion control & To increase  J+H broad-band fringe contrast \\
Active polarization birefringence control & To improve visibility contrast \\
Spatial filtering  & Polarization maintaining single-mode fibers\\
Fiber lengths equalization &  better than $<0.7~$mm \\
Beam combiner & Fiber V-groove supported, all-in-one scheme\\ 
Fringes/photo-metric & 80/20 share for fringes and photometric channels\\
Dispersion & R~=~22, 50, 102, 190 \& 1035 \\
Wollaston and Half-wave plates & For polar-interferometric mode \\
Angular resolution, $\lambda/2B$ & 0.4~mas (J), 0.6~mas (H)\\
Astrometric precision & $<10~\mu$as for close binary\\
H-band sensitivity & H~=~7.5 (R~=~50), H~=~6 (R~=~190)\\
\hline
\end{tabular}
\end{center}
\end{table}

\subsection{MIRC-X observing modes}
Table~\ref{tab:modes} summarizes the list of observing modes available to use and their current commissioning status. MIRC-X has several standard observing modes and a range of experimental observing modes that are under partial commissioning. These experimental modes are shared risk modes, and they require close collaboration with the MIRC-X team as (i) their performance has not yet been fully characterized and/or (ii) the data reduction pipeline is not mature enough to adequately reduce observations conducted in these modes. Although the MIRC-X data reduction pipeline\cite{Anugu2020AJ....160..158A} is mature for the standard observing modes, we encourage PIs to contact or collaborate with the MIRC-X team to utilize available expertise.

Observations of standard MIRC-X modes are, in general straightforward. The observations are typically carried out by the individual PIs who are trained to operate the instrument for science observations. Several PIs already make observations with MIRC-X independently indicating the observing experience with MIRC-X is not complicated. For new PIs, especially from NOIRLab open-access time, the observations are executed by the CHARA Visitor Support Scientist (Gail Schaefer). The observations can be performed on-site and remotely -- using a VNC client maintained by the CHARA Data Scientist (Jeremy Jones). In the remote observing mode, all the real-time operations and commands are executed by the MIRC-X servers that run on the on-site MIRC-X computer. On the remote machine in Atlanta, the observer runs the Linux GTK GUIs that communicate to the server via Secure Shell tunneling using the remote SSH port forwarding protocol.

\begin{table}[ht]
\caption{Available MIRC-X observing modes and commissioning status.}
\label{tab:modes}
\begin{center}       
\begin{tabular}{lll}
\hline
Wavelength/Polarization & Spectral Dispersion & Status and  notes \\
\hline
H & $R$ = 22, 50, 102 \& 190 & Routine. Pipeline ready\cite{Anugu2020AJ....160..158A}. Best sensitivity:\\ 
 & &  H=8.5($R$=22) H=7.5(50), H=6.5(102), H=6(190)\\
 & & Field-of-view: 22mas, 50mas, 102mas \& 190mas, resp. \\
J+H &$R$= 22, 50, 102 \& 190 & Exp. J+H broad-band (lead A. Labdon\cite{Labdon2020arXiv201107865L,Labdon2020_SPIE})\\
H + Wollaston &$R$= 22, 50, 102 \& 190 & Exp. Polarization (lead B. Setterholm\cite{Setterholm2020}) \\
ARMADA for $\lambda$ calib & $R$=50 \& 190 & Exp. For precise astrometry (lead T. Gardner\cite{Gardner2020}) \\ 
High spectral in H &$R$= 1035 & Not tested yet.\\
High spectral in J &$R$= 1077 & Not tested yet.\\
\hline
\end{tabular}
\end{center}
\end{table}

\subsubsection{Standard modes}
H-band observations with spectral dispersion R=22, 50, 102 \& 190 are routine, and the data reduction pipeline is stable and ready to use\cite{Anugu2020AJ....160..158A}. MIRC-X pipeline is written in Python led by J.~B. Le~Bouquin. The code is available at the CHARA Gitlab\footnote{\href{https://gitlab.chara.gsu.edu/lebouquj/mircx\_pipeline}{https://gitlab.chara.gsu.edu/lebouquj/mircx\_pipeline}} and user manual available at the CHARA website\footnote{\href{http://www.chara.gsu.edu/tutorials/mirc-data-reduction}{http://www.chara.gsu.edu/tutorials/mirc-data-reduction}}. 

The selection of spectral resolution is a trade-off between the sensitivity and field-of-view requirements. $R$=50 is the workhorse instrument mode for sensitive observations. The sensitivity of all spectral dispersion units are H=8.5 ($R$=22), H=7.5 ($R$=50), H=6.5 ($R$=102) and H=6 ($R$=190). Higher spectral resolution increases the interferometric field-of-view $\tfrac{R\lambda}{B}$, which is important for observing wide binaries and extended structures. The field-of-views we get for R=22, 50, 102, and 190 are 22~mas, 50~mas, 102~mas, and 190~mas, respectively.

\subsubsection{High spectro-interferometry mode}
The high-resolution spectro-interferometry mode has not yet been tested because of a delay in delivering a new filter-wheel for MIRC-X, which is now planned for installation and testing in the 2021A semester. The $R$ = 1170 for the J band and $R$~=~1035 for the H band grism will enable studies of velocity integrated imaging of the line-emitting regions in J and H-bands. We yet to determine the sensitivity of $R$~=~1035 optic but it is intended for only bright objects as the transmission of this GRISM is low in comparison to the other spectral dispersion units.  We cannot use  J+H together for high spectral resolution measurements because the detector array ($320\times256$~pixels) is not large enough for this.  

\subsubsection{Broad-band J+H wavelength modes}
The broad-band J+H wavelength mode is experimental as it requires active wavelength dispersion control using CHARA Longitudinal Dispersion Corrector (LDC)\cite{Berger2003}. The first simultaneous J+H interferometric observations at $R=50$ are reported in Labdon et al. 2020\cite{Labdon2020arXiv201107865L}. However, the data reduction pipeline is not thoroughly tested. The transmission of LDC optics is sub-optimal at near-infrared wavelengths, reducing the available sensitivity by over a magnitude in J and H-bands.  New LDCs optimized for higher sensitivity are planned to be installed in the following years by the CHARA Array (Lead: T. ten Brummelaar). We plan to make this mode routine in 2021B. 

\subsubsection{Polar-interferometry mode} Polar-interferometry can be used to measure astrophysical polarization signals such as those caused by dust scattering\cite{Ireland2005MNRAS.361..337I}. Polarization observations with a sub-milliarcseconds angular resolution is a mostly unexplored window on the universe. Initial on-sky observations demonstrate hints of detection of polarization signal on YSOs\cite{Setterholm2020}. However, thorough characterization of CHARA's internal polarization is an ongoing project to disentangle the instrumental polarization from the astrophysical signal.   

\subsubsection{ARrangement for Micro-Arcsecond Differential Astrometry (ARMADA)}
The astrometric precision of MIRC-X is $<10~\mu$as in the close-binaries as demonstrated in Anugu et al. 2020\cite{Anugu2020AJ....160..158A} on Iota Peg. However, for the wide binaries, $>100~$mas the astrometric precision is drastically reduced because of inadequate wavelength calibrations: a knowledge of the effective wavelength of the each spectral channel is required and that is limited to $\Delta \lambda/ \lambda = 10^{-3}$ for the MIRC-X standard mode. This limits the astrometric precision to $100\mu$as  for a binary separation of 100mas.  In ARMADA observing mode, the wavelength calibration is improved to $\Delta \lambda/ \lambda = 10^{-4}$ using an Etalon\cite{Anugu2020AJ....160..158A, Gardner2020}. This accurate wavelength calibration enables precise astrometric measurements in the wide binaries. Garder et al. 2020\cite{Gardner2020} report the first results of the ARMADA project. Observations with ARMADA and the data reduction pipeline is currently for experts. We plan to make this mode routine in 2021B.

\section{Initial science results and ongoing science programs}
MIRC-X is making observations since 2017, although the final major engineering was finished in 2019. So far 5 publications came out of the MIRC-X observational data\cite{Kraus2020, Anugu2020AJ....160..158A, Labdon2020arXiv201107865L, Gardner2020, Chiavassa2020}. We highlight a few results in the following sections.

\begin{figure}[ht]
\centering
\includegraphics[width=\textwidth]{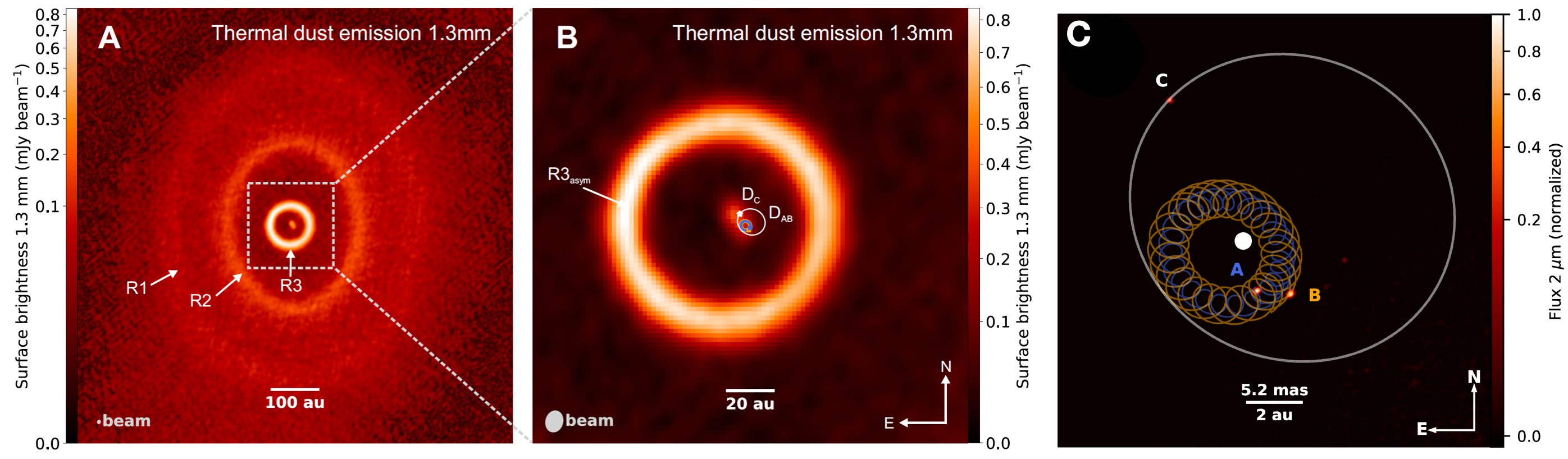}
\caption[example]{\label{Gw_Ori}Kraus et al. 2020\cite{Kraus2020} observed the first disk tearing in a protoplanetary hierarchical triple, GW Ori. (A \& B) ALMA images with 20mas resolution showing disk and misaligned rings. The MIRC-X measured inner triple stellar orbit allowed to link the detected misaligned ring to the disk tearing effect. (C) The MIRC-X reconstructed images of inner triple over-plotted with the best-fitting orbital model. The white dot marks the center-of-mass of the system. Figures reproduced from Kraus et al. 2020\cite{Anugu2020AJ....160..158A}. }
\end{figure}

\subsection{Imaging time-varying structures in protoplanetary disks}
One of our ambitious goals with the MIRC-X is to image highly dynamical processes in the inner regions of protoplanetary disks, e.g., dusty disk winds, variable accretion, and gravitational instabilities. Imaging these time-variable structures in the inner disc regions requires an angular resolution far beyond the reach of conventional telescopes, and the CHARA offered sub-milliarcseconds angular resolutions is the perfect fit for this. The MIRC-X improved sensitivity H=7.5 make a sweet-spot -- increases the number of available YSO targets more than 2 dozen.  

The MIRC-X science team, lead by Stefan Kraus, exploit MIRC-X and CHARA capabilities and execute long-term observations of over 20 YSO targets since 2017 in one or more epochs. The data reduction and aperture synthesis image reconstruction of these targets is an ongoing project -- confident detection of the dynamical process involves a comparison of aperture synthesis images epoch-to-epoch and find time varying structures.

The first refereed paper from MIRC-X observations is a breakthrough result --  the first direct evidence of disk-tearing effect in a protoplanetary hierarchical triple system, GW Ori \cite{Kraus2020}. Hydrodynamic simulations predict the disk-tearing effect for binary systems with significantly inclined disk/orbit planes -- gravitational torque of the binary can tear the disk into separate planes that precess around the inner binary system. The GW Ori observations confirm this theoretical prediction. This detection was made with observations from several facilities, including ALMA, SPHERE, GPI, VLTI, and MIRC-X. The ALMA, SPHERE, and GPI observations show the disk and misaligned rings. The inner triple-star orbit was resolved with optical interferometry measurements collected with MIRC-X and VLTI (see Figure~\ref{Gw_Ori}). 

After this breakthrough detection, the first simultaneous broad-band J+H interferometric observations of a variable YSO, FU~Orionis, were published\cite{Labdon2020arXiv201107865L}. They studied the morphology and temperature gradient of the inner-most regions of the accretion disk, aiming to understand its accretion processes. The MIRC-X observations suggest a temperature gradient power-law derived for the inner disk consistent with theoretical work for steady-state and optically thick accretion disks\cite{Labdon2020arXiv201107865L}. These J+H broad-band observations from MIRC-X are unique as this window is not mainly explored. This provides an opportunity to apply this observing mode on other YSO targets and imaging of stellar surfaces. The J-band window consists of a few interesting lines (e.g., He I $1.08~\mu$m \& Pa-$\gamma$ $1.094~\mu$m). It is an interesting opportunity to explore with the high spectral resolution spectro-interferometry.

\subsection{Companion detection and measuring visual orbits of binary or multiples}

\begin{figure}[ht]
\centering
\includegraphics[width=0.48\textwidth]{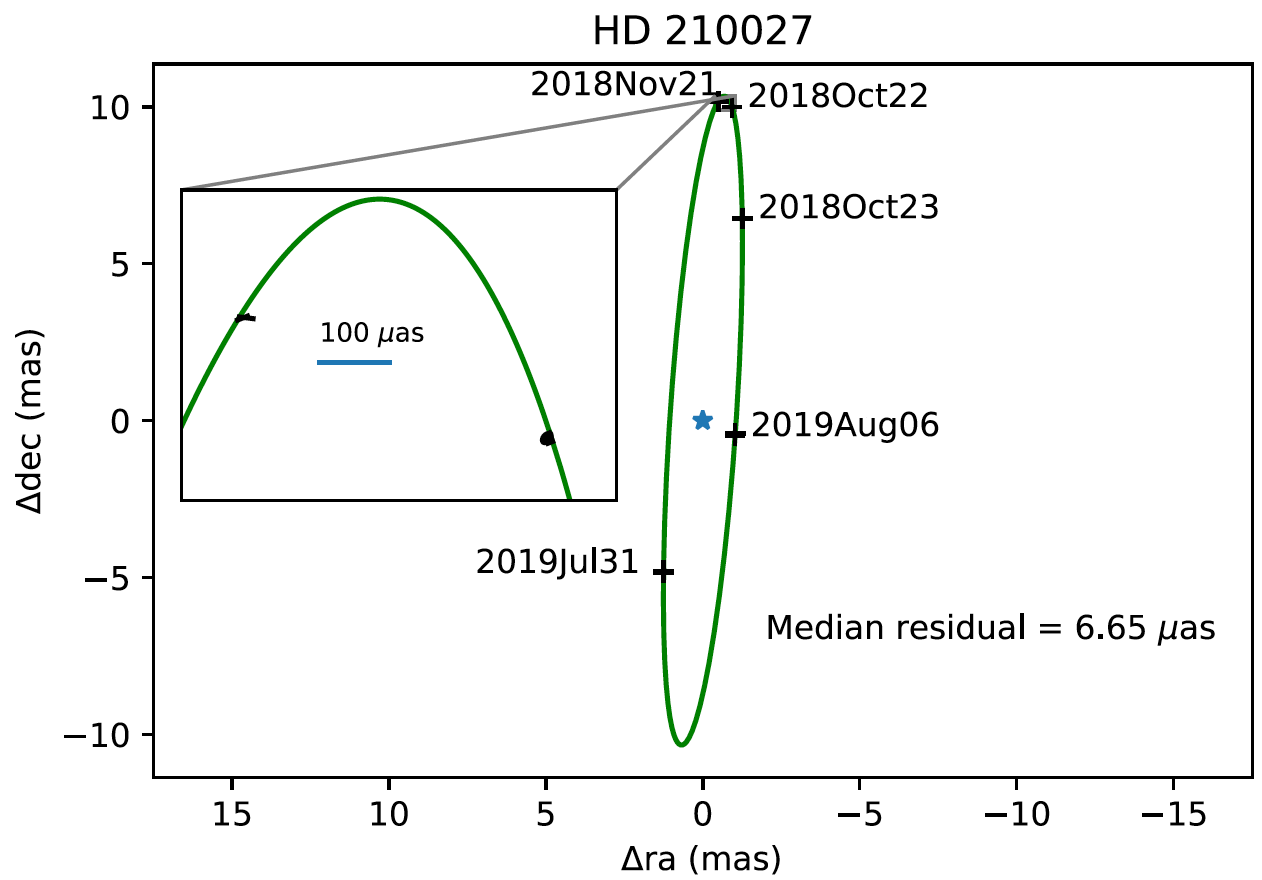}
\includegraphics[width=0.48\textwidth]{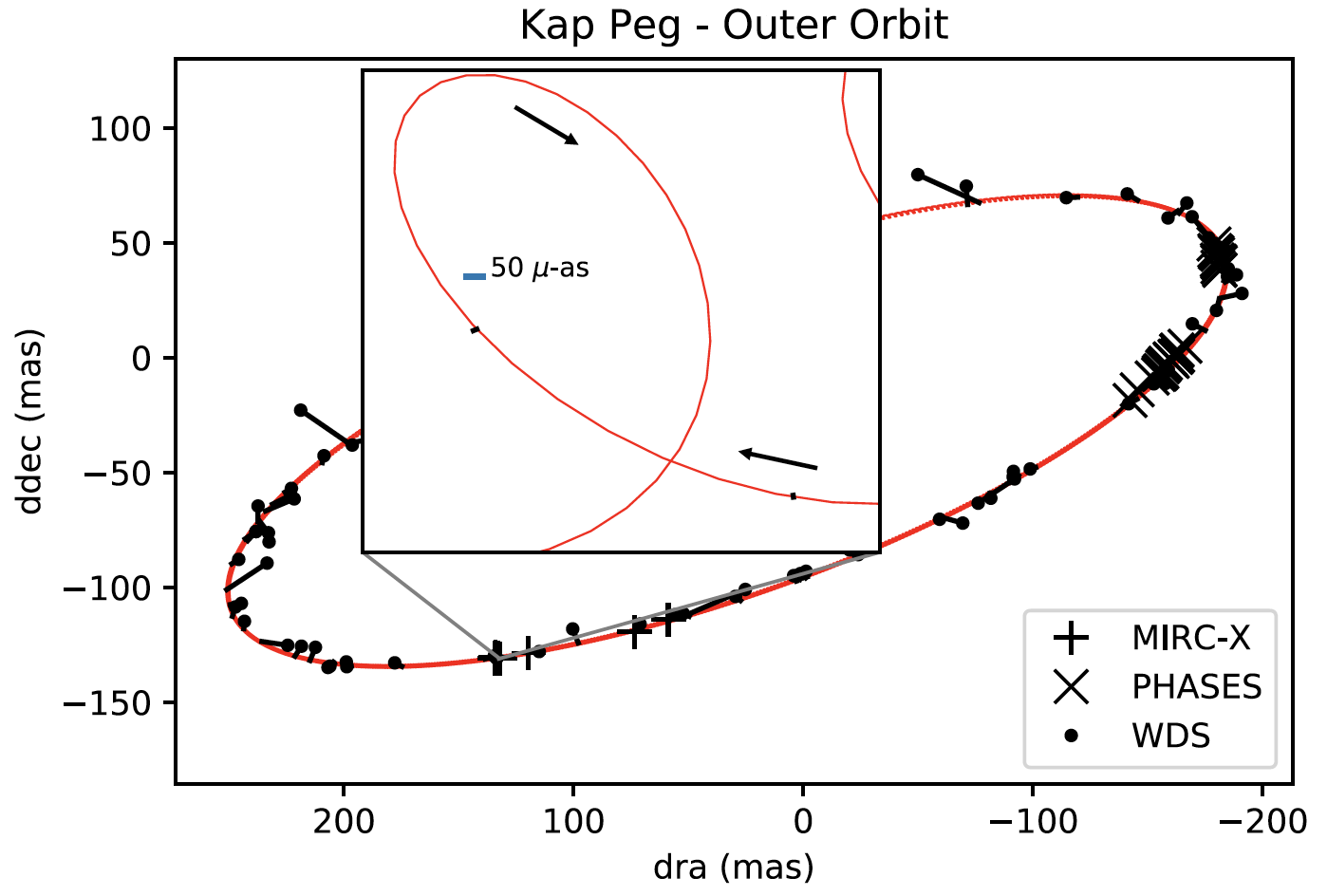}
\caption[example]{\label{Iota_peg} A demonstration of the MIRC-X astrometric precision. (Left) Anugu et al. 2020\cite{Anugu2020AJ....160..158A} show astrometric precision of close binaries -- demonstrating on Iota Peg. The median residual is  $6.65\mu$as in the best orbital fit. (Right) Gardner et al. 2020\cite{Gardner2020} demonstrate an accurate astrometric precision in wide binaries using ARMADA observation mode using $\kappa$ Peg, triple system.}
\end{figure}

\subsubsection{Detecting exoplanets with high-precision astrometry}
The second MIRC-X large-program (lead: T. Gardner and J. Monnier) aims at detecting giant exoplanets and previously unseen low-mass companions orbiting in binary or multiple systems with extremely precise astrometry. The detection of low-mass companions is done by observing the ``wobble'' in the binary orbit caused by the gravity of the unseen companion. This observing technique and survey complement the radial velocity method of detecting exoplanets. The survey selection sample is mainly A or B-type, where the $\sim$au separated giant planet occurrence is difficult with radial velocity methods due to the weak and broad spectral lines of these stars.

\begin{figure}[ht]
\centering
\includegraphics[width=\textwidth]{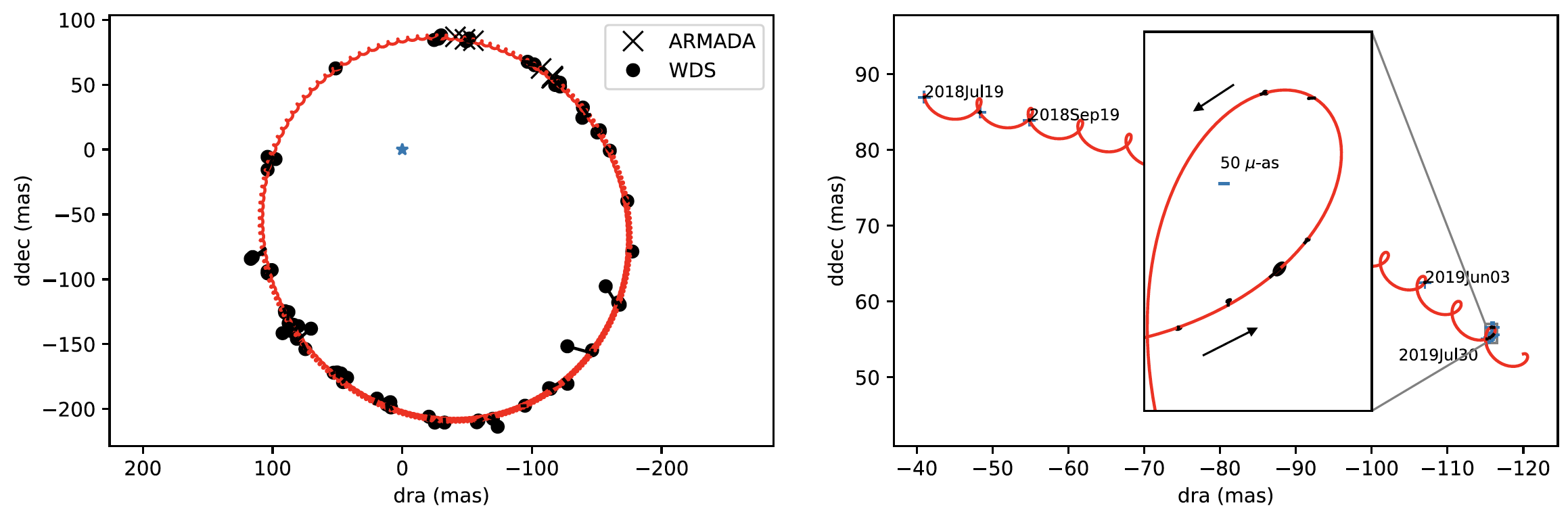}
\caption[example]{\label{alpha_Del} Gardner et al. 2020\cite{Gardner2020} show discovery and measured orbital elements of a 30-day companion to the B-type binary $\alpha$ Del. (Left) The outer binary orbit of $\alpha$ Del as the Ba component moves relative to the A component fixed at the origin. (Right)  Detection of an added astrometric wobble due to the presence of the Bb component. Reproduced from Gardner et al. 2020\cite{Gardner2020}. }
\end{figure}

Gardner et al. 2020\cite{Gardner2020} present the first results (see Figures~\ref{Iota_peg} and \ref{alpha_Del}) of this survey for three targets. These results are very encouraging and capture wobble signatures from the gravitational effects of previously unseen companions. These results are the first steps in forwarding the detection of exoplanets with the MIRC-X ARMADA technique. A few highlights of this work are:

\begin{itemize}
\item Gardner et al. 2020\cite{Gardner2020} validate the MIRC-X ARMADA observing mode  demonstrating precise wavelength calibration to improve systematic errors in binary separation on the $\kappa$~Peg triple system. The MIRC-X ARMADA mode improves the astrometric precision about a factor 10 better in wider binaries separated by $<200$mas. Further improvements require calibration of pupil shifts measurements\cite{Anugu2016SPIE.9907E..27A,Anugu2018a} $<1$mm projected onto the primary telescope mirror space. The pupil shifts will be computed using the Shack-Hartmann pupil images, which are recorded in the telemetry of the CHARA adaptive optics systems\cite{Anugu2020_CHARA_AO,tenBrummelaar2018}. 
    
\item Detected a 30-day companion to a B-type binary $\alpha$ Del. They also measure its orbital elements with the complementary observations from MIRC-X and Radial Velocity observations from the Fairborn Observatory.

\item Measured visual orbit of Be triple star system $\nu$ Gem. This target is interesting for ongoing understanding of multiplicity in Be star systems (Klemet et al., in prep).
\end{itemize}

\subsubsection{Binary detection, orbits, and mass estimations}
In another work, MIRC-X was used to image visual orbits of two Wolf-Rayet stars, WR133 and WR 140,  and then accurately measured their dynamical masses. This work is lead by N.D~Richardson group. They have submitted two papers (Richardson et al., in subm and Thomas et al., in subm) on this project that are in the peer-reviewed stage. Measure of accurate dynamical masses Wolf-Rayet stars is important as the number of Wolf-Rayet stars studied with this sub-milliarcsecond angular resolution level are rare. The accurate mass determination is Wolf-Rayet stars of great importance to the fields of stellar evolution and core-collapse supernovae. This work enables to more accurately predict the masses of black holes that may be formed in core-collapse. 

Other MIRC-X ongoing science programs aimed at detection of new binaries and measuring binary orbits are:
\begin{itemize}
    \item Pre-main-sequence multiple, and other low-mass companion systems provide a unique laboratory to measure the dynamical masses. Those can be used for calibrating the stellar evolutionary models. There is yet a 5\% discrepancy between the fundamental star measured properties and evolutionary models for low mass stars below 0.8 of mass our Sun. A few observing programs with MIRC-X try to address this question by fitting interferometric and spectroscopic orbits and by determining masses of low-mass companions from the binary orbits (Monnier et al. in prep, Kraus et al. in prep,  Schaefer et al., in prep).
    
    \item The interferometric multiplicity survey of O-type stars is very successful\cite{Sana2012Sci...337..444S}. Klemet et al. extend this survey to B-type stars exploiting the MIRC-X capabilities. They want to understand how the multiplicity plays a role in rapidly rotating B-type stars with self-ejected circumstellar disks. They also attempt to study  star-disk connection in Be stars. 
\end{itemize}

\begin{figure}[ht]
\centering
\includegraphics[width=\textwidth]{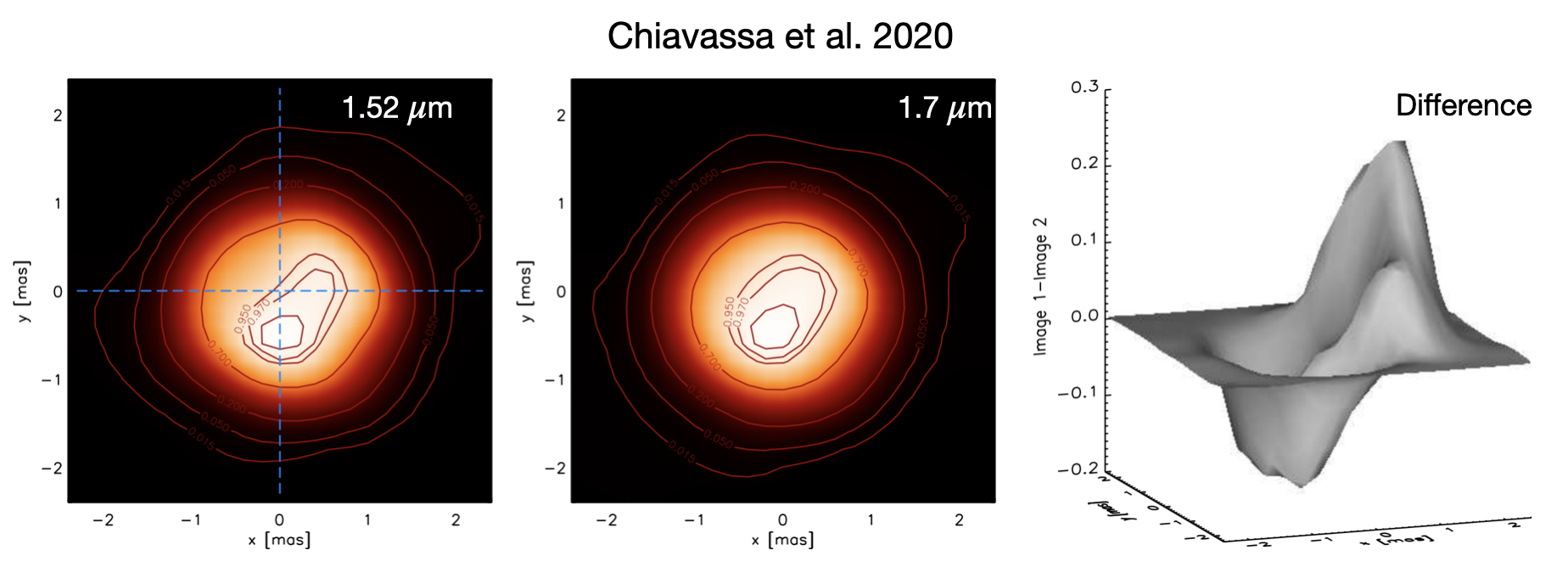}
\caption[example]{\label{Chavissa_AGB} Chavissa et al. 2020\cite{Chiavassa2020} show MIRC-X aperture synthesis imaging of surface of CL Lac, AGB star. The MiRA reconstructed images  at $1.52~\mu$m (left), $1.70~\mu$m (middle) and difference between the two (right). These images show presence of a brighter area that largely affects the position of the photocenter and that likely caused Gaia DR2 astrometric error. Reproduced from Chavissa et al. 2020\cite{Chiavassa2020}.}
\end{figure}

\subsection{Stellar surface imaging}
Stellar surface imaging is one of the main science goals of legacy MIRC instrument, and it has delivered outstanding contributions to the field\cite{Monnier2007, Roettenbacher2016}. We hope to continue this trend with the MIRC-X. 

The first results of MIRC-X are published in Chavissa et al. 2020\cite{Chiavassa2020} by imaging CL Lac (see Figure~\ref{Chavissa_AGB}), Asymptotic Giant Branch star (AGB). This work confirms a correlation between the convection-related variability to its substantial part of the Gaia DR2 parallax error. This result could be used to extract the fundamental properties of AGB stars from the Gaia measurement uncertainties assisted with radiation-hydrodynamic simulations.

A few other ongoing projects further exploit the CHARA sub-milliarcsecond angular resolutions and precision visibility (error $<1$\%) to study stellar surfaces on other systems and these projects are currently in various stages of observing, data reduction, and image reconstruction. 

\begin{itemize}
\item Detection of star spots\cite{Roettenbacher2016} on 
RS Canum Venaticorum variables (Roettenbacher et al., in prep).
\item Imaging convective patterns on the surface of AGB stars as recently illustrated on $\pi^1$ Gruis\cite{Paladini2018} (Abbott et al., in prep).
\item Imaging rapid rotators with MIRC-X as illustrated on Altair\cite{Monnier2007} (Martinez et al., in prep).
\item Imaging of Red-Supergiants\cite{Montarges2018A&A...614A..12M} to understand  relationship between the convection activity and their mass-loss phenomena (Norris et al. prep). Imaging of convection cells on the Betelgeuse surface to understand its recent 2019-2020 unexpected Great Dimming\cite{Dupree2020ApJ...899...68D} and its mass-loss processes (Anugu et al., in prep). The Betelgeuse observations prove the dynamic range of MIRC-X showing calibrated squared visibility down to $V^2=10^{-4}$ very accurately. We got fringes on baselines S1S2 ($B=34~$m), E1E2 ($B=66~$m) and W1W2 ($B=108~$m).
\end{itemize}

\subsection{Post-AGB binary disk imaging}
Post-AGB binaries are unique laboratories to investigate binary stellar evolution. These targets are surrounded by
circumbinary disks that can be seen as second-generation protoplanetary disks\cite{Kluska2020_SPIE, Hillen2016}. Our group exploits the CHARA/MIRC-X capabilities to image the building blocks of six bright post-AGBs available in the Northern sky:  circumbinary disk inner rim, inner binary, accretion signatures. We have observed six targets in multiple epochs to study their binary-disk interactions. Our previous observations show circumbinary disk inner rim, inner binary  and, in addition, complex structures. The detailed image reconstruction of several epochs is in progress (Anugu et al., prep). 

\section{Summary and prospects}
The MIRC-X has proved itself to be a productive astronomical instrument and has been the workhorse instrument for the CHARA Array since its installation in 2017. This trend is expected to continue as CLASSIC/CLIMB\cite{Ten2013} and VEGA\cite{Mourard2009} instruments are on the verge of upgrading.  MIRC-X demonstrated more than two magnitudes of sensitivity improvements compared to legacy MIRC. The CHARA adaptive optics systems are currently under commissioning\cite{tenBrummelaar2018, Anugu2020_CHARA_AO, Schaefer2020}. They will boost sensitivity performance even more, as preliminary results demonstrate promising results with more than a magnitude improvement. All these efforts enable sub-milliarcsecond imaging of more than two dozen faint YSO targets within reach of MIRC-X. Furthermore, this sensitivity boost will push MIRC-X towards the first steps of imaging of Active Galactic Nuclei\cite{GravityCollaboration2020A&A...634A...1G} and direct limb darkening measurements\cite{Kervella2017A&A...597A.137K} of faint JWST candidate exoplanet host targets.  

So far the MIRC-X observing data has produced 5 refereed papers\cite{Anugu2020AJ....160..158A,Labdon2020arXiv201107865L,Gardner2020,Chiavassa2020,Kraus2020}. While we have completed all major engineering for the MIRC-X, we still have several improvements planned for the next semesters:
\begin{itemize}
    \item Efforts are underway to make the MIRC-X data reduction pipeline more robust to work in all practical situations. We are also developing and testing the J+H-band and polarization split observing modes (lead: Labdon and Setterholm). Furthermore, the team is also working on extracting improved differential phases and better calibrated squared visibilities (lead: Monnier; Current square visibility error $<1$\%\cite{Anugu2020AJ....160..158A}). 

    \item Upgrades to the simultaneous J+H-band and polar-interferometric modes are planned for the next semesters to make these observing modes routine and offer to the community. 
    
    \item We will commission filter wheel in 2021 to allow us to use high spectral resolution mode H-band R=1035 and J-band R=1077.  
    
    \item The Observatoire de la C\^{o}te d'Azur (France), in collaboration with the MIRC-X team, developing a software module for fringe tracking mode for MIRC-X that will be used for SPICA instrument, which is expected to be commissioned at the CHARA in 2022\cite{Pannetier2020}.
    
    \item We plan to commission the MIRC-X and MYSTIC simultaneous and coordinated observing in 2021A. The software is already implemented and working at the CHARA as part of the MIRC-X instrument. Two fringe tracking schemes will be tested to enable co-phased operation: (i) \textit{Primary-secondary scheme}\cite{Anugu2020AJ....160..158A} -- one instrument controls the CHARA delay lines and the second one controls the internal delay lines;  Software for this mode is already implemented. (ii) \textit{Weighted scheme} -- for the cases in which ``primary-secondary scheme" is non-optimal. For instance, the cases where zero or very low visibilities depending on baselines length and wavelength. Therefore, based on visibility and SNR, both the instrument's information is combined and weighted before sending the offsets to either the CHARA delay lines or internal delay lines. The software for this observing mode is currently under development. 
\end{itemize}

The MIRC-X design is of the interest to future interferometric designs, e.g., recently proposed Betelgeuse-scope\cite{Anugu2020_SPIE_10.1117/12.2568900}. 

\acknowledgments % equivalent to \section*{ACKNOWLEDGMENTS}       
MIRC-X has been built with support from the European Research Council (ERC) under the European Commission's Horizon 2020 program (Grant Agreement Number 639889). We acknowledge support from the USA National Science Foundation (NSF-ATI 1506540) and NASA-XRP NNX16AD43G.  B.R.S.\ acknowledges support by FINESST: NASA grant \#80NSSC19K1530 and by Michigan Space Grant Consortium: NASA grant \#NNX15AJ20H. A.L.\ acknowledges funding from the UK Science and Technology Facilities Council (STFC) through grant \#630008203. Furthermore, we acknowledge travel funds from STFC PATT grant \#ST/S005293/1.
 This work is based upon observations obtained with the Georgia State University Center for High Angular Resolution Astronomy Array at Mount Wilson Observatory.  The CHARA Array is supported by the National Science Foundation under Grant No. AST-1636624 and AST-1715788.  Institutional support has been provided from the GSU College of Arts and Sciences and the GSU Office of the Vice President for Research and Economic Development. This research has made use of the Jean-Marie Mariotti Center \texttt{Aspro} and \texttt{SearchCal} services.

% References
\bibliography{report} % bibliography data in report.bib
\bibliographystyle{spiebib} % makes bibtex use spiebib.bst

\end{document}